\documentclass[twocolumn,amsmath,amssymb,prd]{revtex4}

\def\al{\alpha}
\def\be{\beta}
\def\ga{\gamma}
\def\de{\delta}
\def\ep{\epsilon}
\def\ve{\varepsilon}

\def\et{\eta}

\def\ka{\kappa}
\def\la{\lambda}

\def\rh{\rho}

\def\si{\sigma}

\def\ph{\phi}

\def\ch{\chi}

\def\om{\omega}

\def\La{\Lambda}

\def\cE{{\cal E}}
\def\cl{{\cal L}}
\def\cL{{\cal L}}

\def\fr#1#2{{{#1} \over {#2}}}
\def\half{{\textstyle{1\over 2}}}
\def\quar{{\textstyle{1\over 4}}}
\def\frac#1#2{{\textstyle{{#1}\over {#2}}}}

\def\vev#1{\langle {#1}\rangle}

\def\lsim{\mathrel{\rlap{\lower4pt\hbox{\hskip1pt$\sim$}}
    \raise1pt\hbox{$<$}}}
\def\gsim{\mathrel{\rlap{\lower4pt\hbox{\hskip1pt$\sim$}}
    \raise1pt\hbox{$>$}}}
\def\sqr#1#2{{\vcenter{\vbox{\hrule height.#2pt
         \hbox{\vrule width.#2pt height#1pt \kern#1pt
         \vrule width.#2pt}
         \hrule height.#2pt}}}}

\def\prt{\partial}

\def\etal{{\it et al.}}

\def\pt#1{\phantom{#1}}

\def\nsc#1#2#3{\om_{#1}^{{\pt{#1}}#2#3}}
\def\lsc#1#2#3{\om_{#1#2#3}}

\def\lulsc#1#2#3{\om_{#1\pt{#2}#3}^{{\pt{#1}}#2}}

\def\vb#1#2{e_{#1}^{{\pt{#1}}#2}}
\def\ivb#1#2{e^{#1}_{{\pt{#1}}#2}}
\def\uvb#1#2{e^{#1#2}}
\def\lvb#1#2{e_{#1#2}}

\newcommand{\beq}{\begin{equation}}
\newcommand{\eeq}{\end{equation}}
\newcommand{\bea}{\begin{eqnarray}}
\newcommand{\eea}{\end{eqnarray}}
\newcommand{\bit}{\begin{itemize}}
\newcommand{\eit}{\end{itemize}}
\newcommand{\rf}[1]{(\ref{#1})}

\begin{document}

\title{
Spontaneous Lorentz Violation, Nambu-Goldstone Modes, and Gravity}

\author{Robert Bluhm$^a$ and V.\ Alan Kosteleck\'y$^b$}

\affiliation{
$^a$Physics Department, Colby College,
Waterville, ME 04901 
\\
$^b$Physics Department, Indiana University,
Bloomington, IN 47405
}

\date{IUHET 478, December 2004}

\begin{abstract}
The fate of the Nambu-Goldstone modes arising from spontaneous
Lorentz violation is investigated.
Using the vierbein formalism,
it is shown that up to 10 
Lorentz and diffeomorphism Nambu-Goldstone modes can appear
and that they are contained within 
the 10 modes of the vierbein associated with gauge degrees
of freedom in a Lorentz-invariant theory.
A general treatment 
of spontaneous local Lorentz and diffeomorphism violation
is given for various spacetimes,
and the fate of the Nambu-Goldstone modes is shown
to depend on both the spacetime geometry and
the dynamics of the tensor field triggering 
the spontaneous Lorentz violation.
The results are illustrated within the general class of
bumblebee models involving vacuum values for a vector field.
In Minkowski and Riemann spacetimes,
the bumblebee model provides a dynamical theory 
generating a photon as a Nambu-Goldstone boson  
for spontaneous Lorentz violation.
The Maxwell and Einstein-Maxwell actions 
are automatically recovered in axial gauge.
Associated effects of potential experimental relevance
include Lorentz-violating couplings 
in the matter and gravitational sectors
of the Standard-Model Extension
and unconventional Lorentz-invariant couplings. 
In Riemann-Cartan spacetime,
the possibility also exists of a Higgs mechanism 
for the spin connection,
leading to the absorption 
of the propagating Nambu-Goldstone modes
into the torsion component of the gravitational field.
\end{abstract}


\maketitle

\section{Introduction}
\label{intro}

Violations of Lorentz symmetry 
arising from new physics at the Planck scale
offer a potentially key signature for
investigations of quantum-gravity phenomenology
\cite{cpt04}.
One elegant possibility is 
that Lorentz symmetry is spontaneously broken 
in an ultimate fundamental theory
\cite{ks}.
The basic idea is that interactions among tensor fields 
in the underlying theory trigger the formation 
of nonzero vacuum expectation values for Lorentz tensors.
The presence of these background quantities throughout spacetime 
implies that Lorentz symmetry is spontaneously broken.

In general,
spontaneous violation of a symmetry has well-established consequences.
The existence of a nontrivial vacuum expectation value
directly modifies the properties of fields that couple to it
and can indirectly modify them through interactions with
other affected fields.
For example,
the vacuum value $\vev{\ph}$ of the Higgs field 
spontaneously breaks the SU(2)$\times$U(1) symmetry 
of the Standard Model (SM),
introducing effective fermion masses 
via couplings of the fermion fields to $\vev{\ph}$.
Moreover,
when a continuous global symmetry is spontaneously broken,
massless modes called Nambu-Goldstone (NG) modes appear 
\cite{ng}.
If instead the local symmetry of a gauge theory is
spontaneously broken,
then the gauge bosons for the broken symmetry 
become massive modes via the Higgs mechanism
\cite{hm}.

For spontaneous Lorentz violation,
the situation is similar.
The existence of a nontrival vacuum value
for a tensor field
affects the behavior of particles coupling to it,
either directly or indirectly through other particles.
These effects can be comprehensively characterized 
in an effective theory for the 
gravitational and SM fields observed in nature
\cite{akgrav,kpo,ck},
via the introduction 
of coefficients for Lorentz violation carrying spacetime indices.
This theory,
called the Standard-Model Extension (SME),
describes the phenomenological implications 
of spontaneous Lorentz violation
independently of the structure of the underlying theory.
To date,
Planck-scale sensitivity has been attained 
to the dominant SME coefficients
in many experiments,
including ones with
photons \cite{photonexpt,photonth},
electrons \cite{eexpt,eexpt2,eexpt3},
protons and neutrons \cite{ccexpt},
mesons \cite{hadronexpt},
muons \cite{muexpt},
neutrinos \cite{nuexpt,nuth},
and the Higgs \cite{higgs},
but a substantial portion of the coefficient space
remains to be explored.

As in the case of internal symmetries,
the vacuum values triggering Lorentz violation 
are accompanied by NG modes
associated with the generators of the broken symmetry
transformations. 
The fate of these NG modes is relevant 
to gravitational and SM phenomenology.
If the massless NG modes are present as such  
and propagate over long distances,
their phenomenology must be compatible with existing
or hypothetical long-range forces.
For example,
it has been suggested that the NG modes 
in a vector theory with spontaneous Lorentz violation 
may be equivalent to electrodynamics in a nonlinear gauge
\cite{dhfb,yn}.
If instead a mass were to develop for the graviton 
in analogy with the usual Higgs mechanism,
other issues would arise.
For example,
even a small mass for the graviton
can modify the predictions of general relativity
and disagree with experiment
\cite{vdvz}.
In fact,
in general relativity with spontaneous Lorentz breaking,
it is known that a conventional Higgs mechanism cannot 
give rise to a mass for the graviton since the analogue of 
the usual Higgs mass term involves derivatives of the metric
\cite{ks2}.
Further complications occur because 
the usual simple counting arguments 
for the number of NG modes can require modification 
in the presence of Lorentz violation
\cite{counting}.

In the context of gravity in a Riemann geometry,
the investigation of spontaneous Lorentz violation was initiated 
with a study of a class of vector theories
\cite{ks2},
called bumblebee models,
that are comparatively simple field theories
in which spontaneous Lorentz violation occurs.
These models and some versions with ghost modes
have since been investigated in a variety of contexts 
\cite{akgrav,kleh,bbmodels}.
There has also been recent interest 
in the timelike diffeomorphism NG mode
that arises when Lorentz symmetry 
is spontaneously broken by a timelike vector.
If such a mode were to appear in a theory
with second-order time derivatives,
it has been shown that it would have an unusual dispersion relation
leading to interesting anomalous spin-dependent forces
\cite{ahclt}.

In this paper,
we investigate the ultimate fate of the NG modes 
associated with spontaneous violation
of local Lorentz and diffeomorphism symmetries.
We perform a generic analysis of theories
formulated in Riemann-Cartan spacetime and its limits,
including the Riemann spacetime of general relativity
and the Minkowski spacetime of special relativity,
and we illustrate the results within the bumblebee model.
The standard vierbein formalism for gravity
\cite{uk}
offers a natural and convenient framework within which to
study the properties of the NG modes,
and we adopt it here.
The basic gravitational fields can be taken as 
the vierbein $\vb \mu a$
and the spin connection $\nsc \mu a b$.
The associated field strengths are the curvature and torsion tensors.
In a general theory of gravity in a Riemann-Cartan spacetime
\cite{rcst},
these fields are independent dynamical quantities.
The usual Riemann spacetime of general relativity is
recovered in the zero-torsion limit,
with the spin connection fixed in terms of the vierbein.
Our focus here is on models in which one or more tensor fields
acquire vacuum values,
a situation that could potentially arise 
in the context of effective field theories 
for a variety of quantum gravity frameworks 
in which mechanisms exist for Lorentz violation.
These include,
for example,
string theory
\cite{ks,kps},
noncommutative field theories
\cite{ncqed},
spacetime-varying fields
\cite{klpe,ahclm,jp},
loop quantum gravity
\cite{qg},
random-dynamics models
\cite{fn},
multiverses
\cite{bj},
and brane-world scenarios
\cite{brane},
so the results obtained in the present work 
are expected to be widely applicable.

The organization of this paper is as follows.
A generic discussion of spontaneous Lorentz violation
in the vierbein formalism is presented in section \ref{slv}.
Section \ref{bbms} discusses basic results for the bumblebee model.
The three subsequent sections, IV, V, and VI,
examine the fate of the NG modes
in Minkowski, Riemann, and Riemann-Cartan spacetimes,
respectively.
Section \ref{concl} contains a summary of the results.
Throughout this work, 
we adopt the notation and conventions 
of Ref.\ \cite{akgrav}.

\section{Spontaneous Lorentz Violation}
\label{slv}

For gravitational theories with a realistic matter sector,
the vierbein formalism
\cite{uk}
is widely used
because it permits a straightforward treatment of fermions
in nontrivial spacetimes.
Since this formalism
distinguishes cleanly between local Lorentz frames
and coordinate frames on the spacetime manifold, 
it is also ideally suited 
for investigations of Lorentz and CPT breaking 
\cite{akgrav},
including the effects of spontaneous violation.

\subsection{General considerations}

A basic object in the formalism
is the vierbein $\vb \mu a$,
which can be viewed as providing 
at each point on the spacetime manifold
a link between the covariant components 
$T_{\la\mu\nu\cdots}$
of a tensor field in a coordinate basis 
and the corresponding covariant components 
$T_{abc\cdots}$
of the tensor field in a local Lorentz frame.
The link is given by 
\beq
T_{\la\mu\nu\cdots} 
= \vb \la a \vb \mu b \vb \nu c \cdots T_{a b c \cdots} .
\label{vier}
\eeq
In the coordinate basis,
the components of the spacetime metric are denoted 
$g_{\mu\nu}$.
In the local Lorentz frame, 
the metric components take the Minkowski form $\et_{ab}$,
but the basis may be anholonomic. 
Expressions 
for contravariant or mixed tensor components 
similar to Eq.\ \rf{vier}
can be obtained by appropriate contractions
with the components $g^{\mu\nu}$ of the inverse spacetime metric. 

The vierbein formalism permits the treatment of 
both basic types of spacetime transformations
relevant for gravitation theories:
local Lorentz transformations, and diffeomorphisms. 
Consider a point $P$ on the spacetime manifold.
Local Lorentz transformations at $P$ act on the tensor components 
$T_{abc\cdots}$
via a transformation matrix $\La^a_{\pt{a}b}$
applied to each index.
For an infinitesimal transformation,
this matrix has the form
\beq
\La^a_{\pt{a}b} \approx \de^a_{\pt{a}b} + \ep^a_{\pt{a}b} ,
\label{LLT}
\eeq
where $\ep_{ab} = - \ep_{ba}$ are the infinitesimal parameters 
carrying the six Lorentz degrees of freedom
and generating the local Lorentz group.
In contrast,
a diffeomorphism is a mapping of $P$ to another point $Q$ 
on the spacetime manifold,
with an associated mapping of tensors at $P$ to tensors at $Q$. 
The pullback of a transformed tensor at $Q$ to $P$ 
differs from the original tensor at $P$.
For infinitesimal diffeomorphisms
characterized in a coordinate basis by the transformation
\beq
x^\mu \rightarrow x^\mu + \xi^\mu ,
\label{diffeos}
\eeq
this difference is given by the Lie derivative
of the tensor $T_{\la\mu\nu\cdots}$
along the vector $\xi^\mu$.
The four infinitesimal parameters $\xi^\mu$ 
comprise the diffeomorphism degrees of freedom.

The vierbein formalism is natural for studies 
of Lorentz violation. 
Spontaneous violation of local Lorentz invariance
occurs when the lagrangian of the theory 
is invariant under local Lorentz transformations
but the vacuum solution violates one or more of the symmetries.
The key feature 
is the existence of a nonzero vacuum expectation value 
for the components $T_{a b c \cdots}$
of a tensor field in a local Lorentz frame 
\cite{akgrav}:
\beq
\vev{T_{a b c \cdots}} \equiv t_{a b c \cdots} \ne 0 .
\label{Tvev}
\eeq 
The values $t_{a b c \cdots}$ 
may be constants or specified functions,
provided they solve the equations of motion of the theory.
Each such expectation value specifies one or more orientations
within any local frame,
which is the characteristic of spontaneous Lorentz violation.

The vacuum expectation value of the vierbein 
is also a constant or a fixed function,
either given by the solution to the gravitational equations
or specified as a background.
For example,
in a spacetime with Minkowski background 
the vacuum value of the vierbein is
$\vev{\vb \mu a} = \de_\mu^{\pt{\mu}a}$
in a suitable coordinate frame.
It follows from Eq.\ \rf{vier} that 
the existence of a vacuum value 
$t_{a b c \cdots}$
for a tensor in a local frame 
implies it also has a vacuum value
$t_{\la \mu \nu \cdots}$ in the coordinate basis
on the manifold.
However,
a nontrivial vacuum expectation value for $t_{\la \mu \nu \cdots}$ 
also implies spontaneous violation of diffeomorphism invariance.
This shows that 
{\it the spontaneous violation of local Lorentz invariance 
implies spontaneous violation of diffeomorphism invariance.} 

In fact,
the converse is also true:
{\it if diffeomorphism invariance is spontaneously broken,
so is local Lorentz invariance.} 
It is immediate that any violation of diffeomorphism invariance 
via vacuum values of vectors or tensors 
breaks local Lorentz invariance,
as above.
An alternative source of diffeomorphism violations 
is possible via vacuum values of scalars
provided the scalars are nonconstant over the spacetime manifold,
but this 
also leads to violations of local Lorentz invariance 
because the derivatives of the scalar vacuum values  
provide an orientation within each local Lorentz frame. 

\subsection{Identification of NG modes}

In discussing the consequences of spacetime-symmetry violations,
it is useful to distinguish among several types of transformations.
Treatments of Lorentz-invariant theories in the literature
commonly define two classes of Lorentz transformations,
called active and passive,
which act on tensor components essentially as inverses of each other.
In a Lorentz-violating theory,
however,
the presence of vacuum expectation values with distinct properties
implies that there are more than 
two possible classes of transformations
\cite{ck}.
For most purposes it suffices 
to limit attention to two possibilities,
called observer transformations
and particle transformations.

{\it Observer} transformations involve changes of the observer frame.
It is standard to assume that any physically meaningful theory 
is covariant under observer transformations,
and this remains true in the presence of Lorentz violation
\cite{ck}.
An observer local Lorentz transformation can be viewed as a rotation
or boost of the basis vectors in the local tangent space.
Tensor components are then expressed in terms of the new basis.
Observer coordinate transformations on the manifold
are general coordinate transformations,
which leave invariant the action.
The statement of observer invariance therefore contains 
no physical information other than the assumption 
of observer independence of the physics.

{\it Particle} transformations are defined to act 
on individual particles or localized fields,
while leaving unchanged vacuum expectation values.
A particle Lorentz transformation
involves a rotation or boost only of localized tensor fields.
The components of the tensor are affected,
while the basis and any vacuum values are unchanged.
Similarly,
particle diffeomorphisms with the pullback incorporated
can be viewed as changes only in localized field distributions,
with the tensor components transforming via the Lie derivative
but the basis and all vacuum values unaffected.

Invariance of a system under particle transformations
has physical consequences,
including notably the existence of conservation laws.
Local Lorentz invariance 
implies a condition on the
antisymmetric components of the energy-momentum tensor $T^{\mu\nu}$,
while diffeomorphism invariance
implies a covariant conservation law for it. 
Thus, 
for example,
in general relativity the laws are
$T^{\mu\nu} = T^{\nu\mu}$
and $D_\mu T^{\mu\nu} = 0$.
Spontaneous breaking of these spacetime symmetries
leaves unaffected the conservation laws.
In contrast,
explicit breaking of these symmetries,
which is described by noninvariant terms in the action,
modifies the laws.
For local Lorentz and diffeomorphism transformations,
the conservation laws in the presence of
spontaneous and explicit breaking are
obtained in Ref.\ \cite{akgrav}
in the context of a general gravitation theory.
 
In a theory with spontaneous breaking of a continuous symmetry,
one or more NG modes are expected.
The NG modes can be identified 
with the virtual excitations around the vacuum solution 
that are generated by the particle transformations 
corresponding to the broken symmetry.
According to the above discussion,
if the extremum of the action involves a nonzero 
vacuum expectation value 
$t_{a b c \cdots}$ 
for a tensor in a local frame,
both local Lorentz invariance 
and diffeomorphism invariance are spontaneously broken.
Since these invariances involve 10 generators
for particle transformations 
\cite{note1,note2},
we conclude that 
{\it up to ten NG modes can appear when an
irreducible Lorentz tensor acquires a vacuum expectation value.}

In the subsequent parts of this work,
it is shown that the vierbein formalism 
is particularly well suited for describing these NG modes.
A simple counting of modes illustrates the key idea. 
The vierbein $\vb \mu a$ has 16 components.
In a Lorentz- and diffeomorphism-invariant theory,
10 of these can be eliminated via gauge transformations,
leaving six potentially physical degrees of freedom 
to describe the gravitational field.
In general relativity,
four of these six are auxiliary and do not propagate, 
leaving only the two usual transverse massless graviton modes;
more general metric gravitational theories 
can have up to six graviton modes
\cite{ellww}.
However,
in a theory with local Lorentz and diffeomorphism violation,
the 10 additional vierbein modes
cannot all be eliminated by gauge transformations
and instead must be treated as dynamical fields in the theory.
In short,
{\it 
the 10 potential NG modes from
spontaneous local Lorentz and diffeomorphism breaking
are contained within the 10 components of the vierbein
that are gauge degrees of freedom 
in the Lorentz-invariant limit.}

\subsection{Perturbative analysis}
\label{Perturbative analysis}

In general,
each NG mode can be obtained by performing on the vacuum
a virtual particle transformation for a broken-symmetry generator 
and then elevating the corresponding spacetime-dependent parameter 
to the NG field.
To identify the NG modes and study their basic properties,
it therefore suffices to consider small excitations about the vacuum 
and to work in a linearized approximation.

If the vacuum solution of a given theory involves the metric 
$g^{\rm vac}_{\mu\nu}$,
then the metric $g_{\mu\nu}$
in the presence of small excitations can be written as
\beq
g_{\mu\nu} = g^{\rm vac}_{\mu\nu} + h_{\mu\nu}.
\eeq
In the general scenario,
distinguishing the background from gravitational fluctuations
requires some care.
For instance, in the shortwave approximation
\cite{ri}
the distinction is made in terms of the amplitude of $h_{\mu\nu}$
and the scales on which $g^{\rm vac}_{\mu\nu}$ and $h_{\mu\nu}$ vary. 
For our purposes,
however,
the presence of a nontrivial background spacetime is unnecessary
and serves to complicate 
the basic study of the properties of the NG modes.
We therefore focus attention here on spacetimes  
in which the vacuum geometry is Minkowski.

Small metric fluctuations about the Minkowski background
can be written as
\beq
g_{\mu\nu} = \et_{\mu\nu} + h_{\mu\nu}.
\eeq
To linear order,
the inverse metric is then 
$g^{\mu\nu} \approx 
\et^{\mu\nu} - \et^{\mu\al} \et^{\nu\be} h_{\al\be}$.
In this context,
the distinction between coordinate indices $\mu$, $\nu$, $\ldots$ 
on the manifold and
the local Lorentz indices $a$, $b$, $\ldots$ 
is diminished,
and Greek letters can be used for both.
The 16-component vierbein can be written as
\beq
\lvb \mu \nu = \et_{\mu\nu} + (\half h_{\mu\nu} + \ch_{\mu\nu}) ,
\label{vhchi}
\eeq
where the ten symmetric excitations $h_{\mu\nu} = h_{\nu\mu}$ are 
associated with the metric,
while the six antisymmetric components $\ch_{\mu\nu} = -\ch_{\nu\mu}$
are the local Lorentz degrees of freedom.

The vacuum expectation value \rf{Tvev} of an arbitrary tensor becomes 
\beq
\vev{T_{\mu\nu\cdots}} \equiv t_{\mu\nu\cdots},
\label{vev2}
\eeq
and excitations about this vacuum value are denoted as 
\beq
\de T_{\la\mu\nu\cdots} = (T_{\la\mu\nu\cdots} - t_{\la\mu\nu\cdots}).
\label{dT}
\eeq
There can be many more such excitations than NG modes.
The NG modes are distinguished by the requirement 
that $\de T_{\la\mu\nu\cdots}$
maintains the extremum of the action
and corresponds to 
broken symmetry generators.

For modes $\de T_{\la\mu\nu\cdots}$ excited via 
local Lorentz transformations or diffeomorphisms,
the magnitude of $T_{\la\mu\nu\cdots}$
at each point is preserved, 
\beq
T^{\la\mu\nu\cdots} g_{\la\al} g_{\mu\be} g_{\nu\ga} 
\ldots T^{\al\be\ga\cdots} = t^2 ,
\label{V}
\eeq
where 
$t^2 = t^{\la\mu\nu\cdots} 
\et_{\la\al} \et_{\mu\be} \et_{\nu\ga} \ldots 
t^{\al\be\ga\cdots}$.
This holds,
for example,
in a theory with potential $V$ 
having the simple functional form 
\beq
V = V(T^{\la\mu\nu\cdots} 
g_{\la\al} g_{\mu\be} g_{\nu\ga} 
\ldots T^{\al\be\ga\cdots} - t^2),
\label{V2}
\eeq
which can trigger a vacuum value $t_{\la\mu\nu\cdots}$ 
when $V$ is extremized.
For instance,
$V$ could be a positive quartic polynomial
in $T^{\la\mu\nu\cdots}$ with minima at zero,
such as $V(x) = \la x^2/2$,
where $\la$ is a coupling constant.
The condition \rf{V} is automatically satisfied by the choice
\beq
T_{\la\mu\nu\cdots} 
= \vb \la \al \vb \mu \be \vb \nu \ga \ldots t_{\al \be \ga \cdots} ,
\label{vier2}
\eeq
which also reduces to the correct vacuum expectation value \rf{vev2}
when the vierbein excitations vanish.
This implies all the excitations in 
$\de T_{\la\mu\nu\cdots}$ 
associated with the NG modes are contained in the vierbein
through Eq.\ \rf{vier2}.

Using the expansion \rf{vhchi} of the vierbein in Eq.\ \rf{vier2}
yields a first-order expression for the tensor excitations 
$\de T_{\la\mu\nu\cdots}$ 
in terms of 
the 16 fields $h_{\mu\nu}$ and $\ch_{\mu\nu}$:
\bea
\de T_{\la\mu\nu\cdots} &\approx &
(\half h_{\la\al} + \ch_{\la\al}) 
t^\al_{\pt{\al}\mu\nu\cdots}
\nonumber \\
&& + (\half h_{\mu\al} + \ch_{\mu\al}) 
t^{\pt{\la}\al}_{\la\pt{\al}\nu\cdots} + \ldots .
\label{Texpansion}
\eea
Evidently,
the combination $(\half h_{\mu\nu} + \ch_{\mu\nu})$ 
contains the interesting dynamical degrees of freedom.

We can observe the effects 
of local Lorentz and diffeomorphism transformations 
by performing each separately.
Under infinitesimal Lorentz transformations,
the vierbein components transform as 
\bea
h_{\mu\nu} 
&\rightarrow& 
h_{\mu\nu} ,
\nonumber\\
\ch_{\mu\nu} &\rightarrow& \ch_{\mu\nu} - \ep_{\mu\nu} ,
\label{LTs}
\eea
while their transformations under infinitesimal diffeomorphisms are
\bea
h_{\mu\nu} 
&\rightarrow& 
h_{\mu\nu} - \prt_\mu \xi_\nu - \prt_\nu \xi_\mu ,
\nonumber\\
\ch_{\mu\nu} 
&\rightarrow& 
\ch_{\mu\nu} - \half (\prt_\mu \xi_\nu - \prt_\nu \xi_\mu ) .
\label{chidiffeo}
\eea
In these expressions,
quantities of order 
$(\ep h)$, $(\ep \ch)$, $(\xi h)$, $(\xi \ch)$, 
etc.\ are assumed small 
and hence negligible in the linearized treatment.

The excitation due to infinitesimal Lorentz transformations is 
\beq
\de T_{\la\mu\nu\cdots} \approx
- \ep_{\la\al} t^\al_{\pt{\al}\mu\nu\cdots}
- \ep_{\mu\al} t^{\pt{\la}\al}_{\la\pt{\al}\nu\cdots} - \ldots .
\label{TLTchange}
\eeq
Depending on the properties of the vacuum value
$t_{\mu\nu\cdots}$,
up to six independent excitations 
associated with broken Lorentz generators
can appear in this expression.
Their nature is determined by 
the six parameters $\ep_{\mu\nu}$.
It follows that the corresponding NG modes $\cE_{\mu\nu}$
for the broken Lorentz symmetries
stem from the antisymmetric components
$\ch_{\mu\nu}$ of the vierbein.

The excitation due to infinitesimal diffeomorphisms is
\bea
\de T_{\la\mu\nu\cdots} & \approx &
- (\prt_\la \xi_\al) t^\al_{\pt{\al}\mu\nu\cdots}
- (\prt_\mu \xi_\al) t^{\pt{\la}\al}_{\la\pt{\al}\nu\cdots} 
- \ldots 
\nonumber\\
&&
- \xi^\al \prt_\al t_{\la\mu\nu\cdots} .
\label{Tdiffchange}
\eea
This can contain up to four independent excitations associated
with broken diffeomorphisms,
depending on the properties of $t_{\mu\nu\cdots}$.
Except for the case of a scalar $T$,
the four potential NG modes $\Xi_\mu$ 
corresponding to the four parameters $\xi_\mu$ 
enter the vierbein accompanied by derivatives.
This can potentially alter their dispersion relations and
couplings to matter currents.

As a simple example,
consider spontaneous breaking due to 
a nonzero vector vacuum value 
$t_\mu$,
which could be timelike, spacelike, or lightlike.
Introduce a quantity
$\cE^\mu = \cE^{\mu\nu} t_{\nu}$
obeying $t_\mu \cE^\nu = 0$.
The three degrees of freedom of $\cE^\mu$ correspond 
to the three Lorentz NG modes 
associated with the three Lorentz generators 
broken by the direction $t_\nu$.
Similarly,
one can introduce a scalar $\Xi^\mu t_\mu$,
which plays the role of the NG mode corresponding
to the diffeomorphism broken by $t_\mu$.
In this example, 
which is studied in more detail in the next section,
there are four potential NG modes.
If a second orthogonal vacuum value $t^\prime_\mu$ is also present,
an additional two Lorentz NG modes appear
because two additional Lorentz generators are broken.
All six Lorentz NG modes $\cE_{\mu\nu}$ enter the theory
once a third orthogonal vacuum value exists.
Similarly, 
as additional vacuum values are added,
more components of the fields $\Xi_\mu$ enter as NG modes
for diffeomorphisms,
until all four are part of the broken theory.

Examples with more complicated tensor representations also
provide insight.
For instance,
consider a theory with an expectation value 
for a two-index symmetric tensor $T^{\mu\nu} = T^{\nu\mu}$.
In this case, 
the choice of vacuum value $t_{\mu\nu}$
can crucially affect the number and type of NG modes.
There is a choice among many possible scenarios.
A subset of the space of possible vacuum values 
consists of those $t_{\mu\nu}$ that
can be made diagonal by a suitable choice of coordinate basis,
but even if attention is restricted to this subset
there are many possibilities. 
For instance,
a vacuum value with diagonal elements $(3,1,1,1)$ 
breaks three boosts and four diffeomorphisms
for a total of seven NG modes,
one with diagonal elements $(4,1,1,2)$
breaks five Lorentz transformations and four diffeomorphisms 
for a total of nine NG modes,
while one with diagonal elements $(6,1,2,3)$
breaks all ten symmetries
for a total of ten NG modes. 

In the general case,
up to ten NG modes can appear
when a tensor acquires a vacuum expectation value 
$t_{\mu\nu\cdots}$.
The fluctuations of the tensor about the vacuum 
under virtual particle transformations are given 
as the sum of the right-hand sides
of Eqs.\ \rf{TLTchange} and \rf{Tdiffchange}.
The associated NG modes consist of 
up to six Lorentz modes $\cE_{\mu\nu}$ and 
up to four diffeomorphism modes $\Xi_\mu$.

The ultimate fate of the NG modes,
and in particular whether 
some or all of them propagate as physical massless fields,
depends on the specific dynamics of the theory.
At the level of the lagrangian,
the linearized approximation involves expanding 
all fields around their vacuum values 
and keeping terms of quadratic order or less.
The dominant terms in the effective lagrangian 
for the NG modes can then be obtained by replacing 
the tensor excitation with the appropriate
NG modes $\cE_{\mu\nu}$ and $\Xi_\mu$
according to Eqs.\ \rf{TLTchange} and \rf{Tdiffchange}.
The resulting dynamics of the modes are
determined by several factors,
including the basic form of the terms in the original action
and the type of spacetime geometry in the theory.
Disentangling these issues 
is the subject of the following sections.

\section{Bumblebee Models}
\label{bbms}

To study the behavior of the NG modes and 
distinguish dynamical effects from geometrical ones,
it is valuable to consider a class of 
comparatively simple models for Lorentz and diffeomorphism violation,
called bumblebee models,
in which a vector field $B^\mu$ 
acquires a constant expectation value $b^\mu$
\cite{akgrav,ks2}.
These models contain many of the interesting features 
of cases with more complicated tensor vacuum values.
For example,
all the basic types of rotation, boost, 
and diffeomorphism violations
can be implemented,
and the existence and properties of the corresponding NG modes 
can be studied for various spacetime geometries.
In this section,
some general results for these models are presented. 

\subsection{Projectors for NG modes}
 
The characteristic feature of bumblebee models
is that a vector field $B^\mu$ acquires a 
vacuum expectation value $b^a$ in a local Lorentz frame.
This breaks three Lorentz transformations and one diffeomorphism,
so there are four potential NG modes.
According to Eq.\ \rf{vier2} and the associated discussion,
the vector field $B^\mu$ can 
be written in terms of the vierbein as 
\beq
B^\mu = \ivb \mu a b^a ,
\label{Bv}
\eeq
which holds in any background metric.
The vierbein degrees of freedom include the NG modes of interest.

As before,
we proceed under the simplifying assumption that the 
background spacetime geometry is Minkowski.
The vacuum solution then takes the form 
\beq
\vev{B^\mu} = b^\mu , \qquad \vev{e_{\mu\nu}} = \et_{\mu\nu} 
\label{vac}
\eeq
in a suitable coordinate frame.
The vierbein can be expanded in terms of
$h_{\mu\nu}$ and $\ch_{\mu\nu}$,
as in Eq.\ \rf{vhchi}.
The fluctuations about the vacuum can therefore be written as 
\beq
\de B^\mu = (B^\mu - b^\mu)
\approx (-\half h^{\mu\nu} + \ch^{\mu\nu}) b_\nu .
\label{name}
\eeq

The results of the previous subsection imply 
that three of the four potential NG modes are contained 
in fields $\cE^\mu$ obeying $b_\mu \cE^\mu = 0$,
while one appears in a combination $\Xi^\mu b_\mu$.
To identify these modes,
it is convenient to separate the excitations \rf{name}
into longitudinal and transverse components 
using projection operators.
Focusing for definiteness on the non-lightlike case ($b^2 \ne 0$),
we define the projectors
\beq
(P_{\parallel})^\mu_{\pt{\mu}\nu} = 
\fr {b^\mu b_\nu} {b^\si b_\si} \, , \quad
(P_{\perp})^\mu_{\pt{\mu}\nu}= 
\de^\mu_{\pt{\mu}\nu} 
- (P_{\parallel})^\mu_{\pt{\mu}\nu} .
\label{projs}
\eeq
The transverse and longitudinal projections of the fluctuations
$\de B^\mu$ can then
be identified as 
\beq
\cE^\mu = (P_{\perp})^\mu_{\pt{\mu}\nu} \de B^\nu
\approx (-\half h^{\mu\nu} + \ch^{\mu\nu}) b_\nu - b^\mu \rh 
\label{epproj}
\eeq
and
\beq
\rh^\mu = (P_{\parallel})^\mu_{\pt{\mu}\nu} \de B^\nu
\approx b^\mu \rh ,
\label{Xiproj}
\eeq
respectively, 
where we have introduced the quantity 
\beq
\rh = - \fr {b^\mu h_{\mu\nu} b^\nu} {2 b^\si b_\si} .
\label{rho}
\eeq
In terms of these fluctuation projections,
the field $B^\mu$ is 
\beq
B^\mu \approx (1+\rh) b^\mu + \cE^\mu .
\label{Bprojs}
\eeq
The reader is warned that at the same level of approximation
the covariant components $B_\mu$ are given by
\beq
B_\mu \equiv  g_{\mu\nu}B^\nu 
\approx (1+\rh) b_\mu + \cE_\mu + h_{\mu\nu}b^\nu .
\eeq

One effect of these projections is to disentangle in $B^\mu$  
the NG modes associated with Lorentz and diffeomorphism breaking.
To see this,
start in the vacuum and perform 
a virtual local particle Lorentz transformation
with parameters $\ep^{\mu\nu}$ 
satisfying $\ep^{\mu\nu} b_\nu \ne 0$.
This leaves unchanged the metric $\et_{\mu\nu}$ 
and the projection $\rh$.
However, 
nonzero transverse excitations $\ep^{\mu\nu} b_\nu$ are generated.
When $-\ep^{\mu\nu}$ is promoted to a field $\cE^{\mu\nu}$,
these become the Lorentz NG modes
$\cE^\mu \equiv \cE^{\mu\nu} b_\nu$.
Note that they automatically obey an axial-gauge condition, 
$b_\mu \cE^\mu =0$,
the significance of which is elaborated in subsequent sections.

Similarly,
the excitations $\rh^\mu$ about the vacuum
contain the diffeomorphism NG mode.
To verify this,
orient the observer coordinate frame so that
the parameter $\xi^\mu$ for the broken diffeomorphism obeys
\beq
\xi^\mu = (P_{\parallel})^\mu_{\pt{\mu}\nu} \xi^\nu
\approx \fr {b^\mu \xi^\nu b_\nu} {b^\si b_\si} ,
\label{xiproj}
\eeq
with $\xi^\mu b^\nu = \xi^\nu b^\mu$.
Then,
a virtual diffeomorphism
generates a nonzero value for $\rh$,
but the field $\cE^\mu$ is unaffected.
Promoting $\xi^\mu$ to the NG field $\Xi^\mu$, 
the expression for $\rh$ becomes 
\beq
\rh = \fr {b^\mu \prt_\mu \Xi_\nu b^\nu} {b^\si b_\si}
= \prt_\mu \Xi^\mu . 
\label{Xirho}
\eeq
We see that 
the longitudinal excitation of $B^\mu$
can indeed be identified with the diffeomorphism NG mode,
as claimed above.
Note that an associated fluctuation in the metric,
given by 
\beq
\et_{\mu\nu} \rightarrow g_{\mu\nu}
\approx \et_{\mu\nu} - \prt_\mu \xi_\nu - \prt_\nu \xi_\mu ,
\label{gdiff}
\eeq
is also generated by the diffeomorphism.

\subsection{Bumblebee dynamics}

The dynamical behavior of $B^\mu$ and the associated NG modes
is determined by the structure of the action
that defines the specific model under study. 
In general,
the lagrangian $\cl_B$
for a single bumblebee field $B^\mu$ 
coupled to gravity and matter
can be written as a sum of terms
\beq
\cL_B = 
\cl_{g} 
+ \cl_{gB}
+ \cl_{K} 
+ \cl_{V}
+ \cl_{J}.
\eeq
Here,
$\cl_{g}$ is the gravitational lagrangian,
$\cl_{gB}$ describes the gravity-bumblebee coupling,
$\cl_{K}$ contains the kinetic terms for $B^\mu$,
$\cl_{V}$ contains the potential, 
including terms triggering the spontaneous Lorentz violation,
and $\cl_{J}$ determines the coupling of $B^\mu$ 
to the matter or other sectors in the model. 

Various forms for each of these partial lagrangians are possible,
and for certain purposes some can be set to zero.
As an explicit example containing all types of terms,
consider the lagrangian
\bea
\cL_B &=& \fr 1 {2 \ka} (e R + \xi e B^\mu B^\nu R_{\mu\nu})
-\frac 1 4 e B_{\mu\nu} B^{\mu\nu} 
\nonumber \\
&& 
- e V (B_\mu B^\mu \pm b^2) - e B_\mu J^\mu ,
\label{BBL}
\eea
where $\ka = 8 \pi G$ and $e \equiv \sqrt{-g}$ is the determinant
of the vierbein.
The Lorentz-invariant limit of this theory 
has been studied previously 
in the context of alternative theories of gravity 
in Riemann spacetime 
\cite{wn}.
The simplified Lorentz-violating limit of the theory with $\xi = 0$
was introduced in 
Ref.\ \cite{ks},
while the theory \rf{BBL} and some versions 
including ghost vectors have been explored more recently in 
Refs.\ \cite{akgrav,kleh,bbmodels}.

In the model \rf{BBL} and its subsets,
which are among those used in the sections below, 
the gravitational lagrangian $\cl_{g}$ is that of general relativity.
The specific nonminimal gravity-bumblebee interaction $\cl_{gB}$ 
in this example is controlled by the coupling constant $\xi$.
The last term in Eq.\ \rf{BBL}
represents a matter-bumblebee interaction $\cl_{J}$,
involving the matter current $J^\mu$.
The theory \rf{BBL} is written for a
Riemann or Riemann-Cartan spacetime,
but the limit of Minkowski spacetime is also of interest below,
and the corresponding lagrangian can be obtained 
by eliminating the first two terms and setting $e=1$.

The partial lagrangian $\cl_{V}$ in the model \rf{BBL}
involves a potential $V$ of the general form in Eq.\ \rf{V2},
inducing the spontaneous Lorentz and diffeomorphism violation.
The quantity $b^2$ is a real positive constant,
related to the vacuum value $b^a$ of the bumblebee field 
by $b^2 = |b^a \et_{ab} b^b|$,
while the $\mp$ sign in $V$ determines whether 
$b^a$ is timelike or spacelike.
As before, 
$V$ can be a polynomial such as $V(x) = \la x^2/2$, 
where $\la$ is a coupling constant.
An alternative explicit form for $V$ of particular value 
for studies of NG modes is the sigma-model potential
\beq
V (B_\mu B^\mu \pm b^2) = \la (B_\mu B^\mu \pm b^2) ,
\label{Vsm}
\eeq
where the quantity $\la$ is now a Lagrange-multiplier field.
The Lagrange multiplier acts to constrain the theory 
to the extrema of $V$ obeying $B_\mu B^\mu \pm b^2 = 0$,
thereby eliminating fields other than the NG modes.
This model is a limiting case of the previous polynomial one
in which the massive mode is frozen. 
Note that in any case the potential $V$  
excludes the possibility of a $U(1)$ gauge invariance
involving $B^\mu$, 
whatever the form of the bumblebee kinetic term in the action. 

The kinetic partial lagrangian $\cl_{K}$ in the model of Eq.\ \rf{BBL}
involves a field strength $B_{\mu\nu}$ for $B_\mu$,
defined as
\beq
B_{\mu\nu} = D_\mu B_\nu - D_\nu B_\mu ,
\label{Bmunu}
\eeq
where $D_\mu$ are covariant derivatives
appropriate for the chosen spacetime geometry.
In a Riemann or Minkowski spacetime,
where the torsion vanishes,
this field strength reduces to
$B_{\mu\nu} = \prt_\mu B_\nu - \prt_\nu B_\mu$. 
In either case,
with $B^\mu$ given by Eq.\ \rf{Bprojs},
it follows that the diffeomorphism mode 
contained in $\rh$ cancels in 
the kinetic partial lagrangian $\cL_K$
in Eq.\ \rf{BBL},
and $\cL_K$ propagates two modes.
If an alternative kinetic partial lagrangian $\cl_{K}$ 
is adopted instead
an additional mode can propagate.
A simple example is given by the choice 
\cite{jm}
\beq
\cL_{K,\rm ghost} = \half e B_\mu D^\nu D_\nu B^\mu ,
\label{ghke}
\eeq
which propagates three modes,
although this kinetic partial lagrangian can be unphysical as such 
because it can contain ghost dynamics.

The above discussion demonstrates that
the fate of the Lorentz and diffeomorphism NG modes 
depends on the geometry of the spacetime
and on the dynamics of the theory.
To gain further insight into the nature and fate of the NG modes,
we consider in turn the three cases of  
Minkowski, Riemann, and Riemann-Cartan spacetimes,
each in a separate section below.

\section{Minkowski Spacetime}
\label{flat}

\subsection{Role of the vierbein}

In Minkowski spacetime,
the curvature and torsion vanish by definition, 
and global coordinate systems exist such that
\beq
g_{\mu\nu} = \et_{\mu\nu}.
\label{etg}
\eeq
Particle Lorentz transformations can be performed using
\beq
\La_\mu^{\pt{\mu}\nu} \approx 
\de_\mu^{\pt{\mu}\nu} + 
\ep_\mu^{\pt{\mu}\nu}
\label{plt}
\eeq
to rotate and boost tensor components.
These transformations are global if 
$\ep_\mu^{\pt{\mu}\nu}$ 
is independent of the spacetime point;
otherwise, they are local. 
In any case,
they maintain the metric 
in the form $\et_{\mu\nu}$ at each point.
It is instructive to compare the expression \rf{plt} 
with the form of the vierbein when $h_{\mu\nu} = 0$:
\beq
\vb \mu \nu \approx 
\de_\mu^{\pt{\mu}\nu} 
+ \ch_\mu^{\pt{\mu}\nu}.
\label{vbmk}
\eeq
It follows that
a vierbein with $h_{\mu\nu} = 0$ in Minkowski spacetime 
can be identified with a local particle Lorentz transformation.
Also,
starting from a vacuum solution with
$\vb \mu \nu = \de_\mu^{\pt{\mu}\nu}$,
a local Lorentz transformation \rf{plt} generates
$\ch_\mu^{\pt{\mu}\nu}= \ep_\mu^{\pt{\mu}\nu}$.

In Minkowski spacetime,
the diffeomorphisms maintaining $h_{\mu\nu} = 0$
are global spacetime translations,
corresponding to constant $\xi^\mu$
in cartesian coordinates. 
Under local diffeomorphisms with nonconstant $\xi^\mu$,
the metric transforms as 
\cite{note3}
\beq
\et_{\mu\nu} \rightarrow g_{\mu\nu} = \et_{\mu\nu} + h_{\mu\nu},
\label{metricmk}
\eeq
where 
$h_{\mu\nu} = - \prt_\mu \xi_\nu - \prt_\nu \xi_\mu$. 
A corresponding term is generated in the vierbein,
given by Eq.\ \rf{chidiffeo}.

To study the NG modes in Minkowski spacetime,
consider first a theory with tensor field $T_{\la\mu\nu\cdots}$ 
satisfying the condition \rf{V}.
As before,
the NG modes can be identified with small excitations 
$\de T_{\la\mu\nu\cdots}$ 
maintaining this constraint.
The constraint is invariant 
under global Lorentz transformations and translations,
as is evident by writing Eq.\ \rf{V}
in a coordinate frame in which Eq.\ \rf{etg} holds.
More importantly,
the constraint is also invariant under 
local Lorentz transformations and diffeomorphisms.
It follows that
even in Minkowski spacetime there are up to ten NG modes
when spontaneous Lorentz violation occurs.

As a more explicit example,
consider a bumblebee model
with the vacuum solution of the form \rf{vac}.
Local Lorentz transformations and local diffeomorphisms 
change the components $B^\mu$ and the metric $g_{\mu\nu}$
while maintaining the equality $B^\mu g_{\mu\nu} B^\nu = \mp b^2$,
but some of these transformations 
are spontaneously broken by the vacuum values.
As before,
there are potentially four NG modes that can appear,
consisting of three Lorentz NG modes
$\cE^\mu \equiv \cE^{\mu\nu} b_\nu$ obeying $b_\mu \cE^\mu =0$,
and one diffeomorphism NG mode contained in $\rh$ 
and given by Eq.\ \rf{Xirho}.

In a bumblebee model,
the vacuum solution can break global Lorentz transformations 
while preserving translations,
so that energy-momentum is conserved.
This type of assumption is sometimes adopted 
within the broader context of the SME 
as a useful simplification for experimental studies. 
Translation symmetry holds if $b^\mu$ is a constant in 
a coordinate frame in which the metric takes the form \rf{etg}.
However,
the vacuum solution $b^\mu$ in this frame could in principle 
also be a smoothly varying function of spacetime obeying 
$b^\mu \et_{\mu\nu} b^\nu = \mp b^2$,
such as a soliton solution.
Then,
$b^\mu$ has constant magnitude 
but different orientation at different spacetime points,
and both global Lorentz and translation symmetries are broken.
In this case,
a vierbein can be introduced at each point that 
transforms $b_\mu$ into a local field $b_a$ 
obeying $b_a b^a = \mp b^2$ at each point
but having components that are constant over the spacetime.
The role of the vierbein in this context can be regarded
as a link to a convenient anholonomic basis 
in which $b_a$ appears constant. 
Whatever the fate of the global transformations,
however,
local Lorentz transformations and diffeomorphisms 
are broken by the vacuum solution \rf{vac},
and the behavior of the four potential NG modes 
contained in $\cE^\mu$ and $\rh$
is determined by the dynamics of the theory. 

\subsection{Fate of the NG modes}

Since gravitational excitations are absent in Minkowski spacetime,
no kinetic terms for $h_{\mu\nu}$ can appear
and there is no associated dynamics.
Any propagation of NG modes must therefore
originate from lagrangian terms involving 
$T^{\la\mu\nu\cdots}$.
Diffeomorphisms produce infinitesimal excitations
of the vacuum solution given by \rf{Tdiffchange},
which generate NG modes in the combination 
$\prt_\mu \Xi_\al$.
It might therefore seem that even nonderivative terms 
for $T^{\la\mu\nu\cdots}$ in the lagrangian 
could generate derivative terms for some NG modes
and hence possibly lead to their propagation.
However,
when a potential $V$ drives the breaking,
any nonderivative term in $T^{\la\mu\nu\cdots}$ 
is intrinsically part of $V$,
so its presence may affect 
the specific form of the vacuum solution \rf{vev2}
but cannot contribute to the propagator for the NG modes.
Indeed, 
no contributions from $V$ arise in 
the effective action for the NG modes
because this action is obtained 
via virtual particle transformations 
leaving $V$ invariant and at its extremum.
This result can equivalently be obtained
using the vierbein decomposition 
\rf{Texpansion} of $T^{\la\mu\nu\cdots}$,
since this expansion automatically extremizes $V$
and also contains the NG modes as shown before.
It follows in this case that any propagation of NG modes
must be determined by kinetic or derivative-coupling terms 
for $T^{\la\mu\nu\cdots}$.

Next,
we illustrate some of the possibilities 
for generating propagators for the NG modes
using kinetic terms
in a bumblebee model.
Consider first a special Minkowski-spacetime limit 
of the theory \rf{BBL},
for which the lagrangian is
\beq
\cL_B = 
-\frac 1 4 B_{\mu\nu} B^{\mu\nu} 
- \la (B_\mu B^\mu \pm b^2) - B_\mu J^\mu .
\label{flatBB}
\eeq
The ghost-free kinetic term chosen involves 
the zero-torsion limit of the field strength $B_{\mu\nu}$
in Eq.\ \rf{Bmunu},
and the adoption of the sigma-model potential \rf{Vsm}
ensures a focus on the NG modes.

For this theory,
a coordinate frame can be chosen in which the vacuum solution is 
\beq
\vev{B^\mu} = b^\mu , \qquad
\vev{\lvb \mu \nu} = \et_{\mu\nu} , \qquad \vev{\la} = 0 .
\label{flatBBvac}
\eeq
For simplicity,
in what follows we take $b^\mu$ to be constant in this frame.
The relevant virtual fluctuations of the bumblebee field 
around the vacuum solution,
generated by the broken particle Lorentz transformations 
and diffeomorphisms,
can be decomposed using the projector method.
The result is Eq.\ \rf{Bprojs},
where $\cE_{\mu}$ and $\rh$ contain the NG modes.
As before,
the Lorentz NG modes satisfy
an axial-gauge condition $b_\mu \cE^\mu = 0$
and so represent three degrees of freedom.

With the vacuum \rf{flatBBvac} and
the choice of kinetic term in Eq.\ \rf{flatBB}, 
the diffeomorphism mode contained in $\rh$ cancels in $B_{\mu\nu}$.
It therefore cannot propagate
and is an auxiliary mode.
In fact, 
the kinetic term in Eq.\ \rf{flatBB} reduces 
to the form of a U(1) gauge theory in an axial gauge,
so one of the three Lorentz modes is auxiliary too.
Adopting the suggestive notation 
$\cE_\mu \equiv A_\mu$ 
and denoting the corresponding field strength by 
$F_{\mu\nu} \equiv \prt_\mu A_\nu - \prt_\nu A_\mu$,
we find that the lagrangian \rf{flatBB}
reduces at leading order to  
\beq
\cL_B \to \cL_{\rm NG} \approx 
-\frac 1 4 F_{\mu\nu} F^{\mu\nu} - A_\mu J^\mu 
- b_\mu J^\mu + b^\mu \prt_\nu \Xi_\mu J^\nu .
\label{linflatBB}
\eeq
This is the effective quadratic lagrangian 
determining the propagators of the NG modes
in the theory \rf{flatBB}.
Note that the axial-gauge condition $b_\mu A^\mu = 0$
includes the special cases of temporal gauge 
($A^0 = 0$) 
and pure axial gauge
($A^3 = 0$),
and it ensures the constraint term is absent in $\cL_{\rm NG}$. 
Note also that varying with respect to $\Xi_\mu$
yields the current-conservation law, 
$\prt_\mu J^\mu = 0$. 

We see that the NG modes for the Minkowski-spacetime bumblebee
theory \rf{flatBB}
have the basic properties of the massless photon,
described as a U(1) gauge theory in an axial gauge.
This result is consistent with an early analysis by Nambu
\cite{yn},
who investigated the constraint $B_\mu B^\mu = \mp b^2$
as a nonlinear gauge choice 
that spontaneously breaks Lorentz invariance.
In the linearized limit with $B^\mu \approx b^\mu + \cE^\mu$,
this gauge choice reduces to an axial-gauge condition
$b_\mu \cE^\mu = 0$ at leading order.
The discussion here involves a Lagrange-multiplier constraint
rather than a direct gauge choice 
and so the theories differ,
but the result remains unaffected. 

The masslessness of the photon 
in the effective theory \rf{linflatBB}
is a consequence of the spontaneously broken Lorentz symmetry
in the original 
theory \rf{flatBB}.
An interesting question 
is whether this idea has experimentally verifiable consequences.
Indeed,
a version of the theory \rf{flatBB} with an explicit matter sector
has been presented in Ref.\ 
\cite{kleh}
as a model for quantum electrodynamics (QED)
that generates a Lorentz-violating term in the SME limit.
The latter appears in Eq.\ \rf{linflatBB}
as the Lorentz-violating term $b_\mu J^\mu$,
along with a conventional charged-current 
interaction $A_\mu J^\mu$.

If the current $J^\mu$ represents 
the usual electron current in QED,
then the term $b_\mu$
can be identified with the coefficient $a^e_\mu$ 
for Lorentz violation in the QED limit of the SME
\cite{ck}.
If this coefficient is spacetime independent,
it is known to be unobservable in experiments
restricted to the electron sector
\cite{bekcm},
but coefficients of this type
can generate signals in the quark 
\cite{hadronexpt,bckp}
and neutrino
\cite{ck,nuth}
sectors.
Moreover,
various other possible sources of experimental signals
can be considered, 
such as spacetime dependence of the coefficients,
the presence of multiple fields and other types of current,
and interference between several sources of Lorentz violation.
Nonminimal couplings to other sectors,
including the gravitational couplings discussed in the next section,
can also produce experimental signals.
All these effects are contained within the SME.
More radical options for interpretation of the NG modes
from Lorentz and diffeomorphism breaking
in a general theory can also be envisaged,
ranging from new long-range forces weakly coupled to matter 
with possible implications for dark matter and dark energy
to the identification of many or all massless modes in nature
with the NG modes.
A careful investigation of these possibilities 
lies outside our present scope
but would be of definite interest. 

Another interesting issue is whether 
a consistent theory exists in Minkowski spacetime 
in which the diffeomorphism mode contained in $\rh$ propagates. 
Consider, 
for example,
substituting an alternative kinetic term 
of the form \rf{ghke} in the lagrangian \rf{flatBB},
yielding the model
\beq
\cL_{B, \rm ghost} = \half B_\mu \prt^\nu \prt_\nu B^\mu 
- \la (B_\mu B^\mu \pm b^2) - B_\mu J^\mu 
\label{toyBB}
\eeq
in cartesian coordinates.
This model may have a ghost,
but the behavior of the NG modes can nonetheless be examined.
Proceeding via the projector method as before,
we find the kinetic term becomes 
$\half \cE^\mu \prt^\nu \prt_\nu \cE_\mu$,
so the diffeomorphism NG mode 
contained in $\rh$ is auxiliary
while the three $\cE_\mu$ modes propagate.
More generally,
in the covariant derivative $D_\mu B_\nu$ 
relevant for a general coordinate system in Minkowski spacetime,
the NG excitations reduce to $\prt_\mu \cE_\nu$
when $b_\nu$ is constant in cartesian coordinates,
so kinetic terms contain no propagation
of the diffeomorphism mode $\Xi_\mu$ in this case.

This example illustrates a general difficulty
in forming a covariant kinetic term 
that permits propagation of the diffeomorphism modes 
for the case of constant vacuum value $t_{\la\mu\nu\cdots}$.
To be observer independent,
the Minkowski-spacetime lagrangian must be formed 
from contractions of a tensor $T^{\la\mu\nu\cdots}$
and its derivatives.
Only terms with one or more derivatives can contribute
to the propagation,
as shown above. 
However,
for covariant derivatives of $T^{\la\mu\nu\cdots}$
with constant $t_{\la\mu\nu\cdots}$,
the diffeomorphism modes always enter combined with a derivative,
$\prt_\mu \Xi_\al$,
while the connection
acquires a corresponding change 
induced by the metric fluctuation \rf{gdiff}
that cancels them.
Other possibilities 
would therefore need to be countenanced,
such as a nonconstant $t_{\la\mu\nu\cdots}$.
In any case,
the structure of terms 
containing derivatives of $\Xi_\mu$ 
in the effective lagrangian for the NG modes 
represents a major difference between 
spontaneous breaking of internal and spacetime symmetries.
In the former,
the relevant fields carry internal indices 
that are independent of spacetime derivatives,
and so theories with compact internal symmetry groups can
be constructed that propagate all the NG modes 
without generating ghosts.
In contrast,
the spontaneous violation of spacetime symmetries 
involves fields with spacetime indices,
and for the diffeomorphism NG modes this 
changes the derivative structure 
in the effective lagrangian.

\section{Riemann Spacetime}
\label{riemann}

In this section,
we revisit for Riemann spacetimes
the results obtained in the Minkowski spacetime case.
The general features obtained in the previous section 
apply to a nondynamical Riemann spacetime 
with fixed background metric.
The primary interest here therefore lies instead with 
Riemann spacetimes having a dynamical metric. 

\subsection{Vierbein and spin connection}

In a Riemann spacetime with dynamical metric $g_{\mu\nu}$,
the nature and properties of the NG modes 
for spontaneous Lorentz and diffeomorphism violation
can still be analyzed following the general approach 
of sections \ref{slv} and \ref{bbms}.
We assume that the solutions to a theory for a tensor 
$T_{\la\mu\nu\cdots}$ 
satisfy the condition \rf{V},
thereby inducing a nonzero vacuum value 
$t_{\la\mu\nu\cdots}$.
This condition is automatically satisfied 
by writing $T_{\la\mu\nu\cdots}$ in terms of the vierbein,
\beq
T_{\la\mu\nu\cdots}
= \vb \la a \vb \mu b \vb \nu c \ldots t_{a b c \cdots} ,
\label{vier3}
\eeq
where
$t_{a b c \cdots}$ is the vacuum value of the tensor 
in a local Lorentz frame.
For definiteness in what follows,
we suppose that $t_{a b c \cdots}$ 
is constant over the spacetime manifold in the region of interest.
This assumes appropriate smoothness of $t_{\la\mu\nu \cdots}$
and compatibility with any boundary conditions.
For example,
if the Riemann spacetime is asymptotically Minkowski
and the vacuum value of the vierbein $e_{\mu\nu}$ 
is taken as $\et_{\mu\nu}$,
then the components of $t_{\la\mu\nu \cdots}$ must be 
asymptotically constant.

A primary difference in Riemann spacetime is that 
up to six of the 16 independent components of the vierbein $\vb \mu a$
can represent dynamical degrees of freedom of the gravitational field.
The lagrangian for the theory 
must therefore contain dynamical terms for the vierbein.
This raises the issue of the effect of these additional terms
on the other 10 components of the vierbein,
all of which are potential NG modes
for the spontaneous violation of spacetime symmetries.

At first sight
the situation might seem to be further complicated by the existence 
of the spin connection 
$\nsc \mu a b$,
which permits the construction of the covariant derivative
and in principle can have up to 24 independent components.
However,
the requirement that the connection be metric,
\beq
D_\la \vb \mu a = 0,
\eeq
and the vanishing of the torsion tensor
in a Riemann spacetime
imply that the spin connection $\nsc \mu a b$ 
can be specified completely 
in terms of the vierbein and its derivatives according to
\bea
\nsc \mu a b &=&
\half \uvb \nu a ( \prt_\mu \vb \nu b - \prt_\nu \vb \mu b)
- \half \uvb \nu b ( \prt_\mu \vb \nu a - \prt_\nu \vb \mu a)
\nonumber\\
&&
- \half \uvb \al a \uvb \be b \vb \mu c
(\prt_\al \lvb \be c - \prt_\be \lvb \al c).
\label{scvierb}
\eea
It is therefore sufficient to consider the
behavior of the vierbein 
in studying the properties of the NG modes. 
For example,
a Higgs mechanism is excluded for the spin connection 
in a Riemann spacetime,
since no independent propagating massless modes for $\nsc \mu a b$ 
exist to absorb the NG degrees of freedom.
In the next section,
we revisit this issue in the context 
of the more general Riemann-Cartan geometry,
for which the spin connection is an independent dynamical field.

Covariant derivatives acting on  
$T_{\la\mu\nu\cdots}$ 
in the lagrangian can also generate
propagators for the vierbein and hence for the NG modes.
The covariant derivative $D_\al T_{\la\mu\nu\cdots}$ 
is given by
\beq
D_\al T_{\la\mu\nu\cdots} 
= \vb \la a \vb \mu b \vb \nu c \ldots D_\al 
t_{a b c \cdots} .
\label{Dvier3}
\eeq
The term $D_\al t_{a b c \cdots}$ in this equation 
contains products of the spin connection with
the vacuum value $t_{a b c \cdots}$,
which according to Eq.\ \rf{scvierb}
generates expressions involving a single derivative of the vierbein.
In the presence of spontaneous violation of spacetime symmetries,
it follows that any piece of the lagrangian
involving a quadratic power of $D_\al T_{\la\mu\nu\cdots}$ 
can produce quadratic-derivative terms involving the vierbein.

The above discussion shows that 
in a Riemann spacetime the fate of the NG modes 
can depend on both the gravitational terms in the lagrangian
and the kinetic or other derivative-coupling terms 
for the tensor field.
In what follows,
we consider implications of these results
for bumblebee models.

\subsection{Bumblebee and photon}

For a bumblebee model
in an asymptotically flat Riemann spacetime,
the vacuum structure is similar to the Minkowski case.
We take the vacuum values for $B^\mu$ and the vierbein
to be those of Eq.\ \rf{vac}.
The projector method can be applied,
leading to the decomposition \rf{Bprojs} of $B^\mu$.
There are four potential NG modes
contained in the fields $\cE^\mu$ and $\rh$,
and the axial-gauge condition $b_\mu \cE^\mu = 0$ holds.
Note that the field strength $B_{\mu\nu}$ in Eq.\ \rf{Bmunu}
can be rewritten to give
\beq
B_{\mu\nu} = (\prt_\mu \vb \nu a - \prt_\nu \vb \mu a) b_a ,
\label{vB}
\eeq
where $b_a$ is taken constant
as in the previous subsection.

The properties of the NG modes depend on 
the kinetic terms for $B^\mu$ 
and the gravitational terms in the lagrangian.
To gain further insight,
consider the lagrangian \rf{BBL} 
with a Lagrange-multiplier potential,
\bea
\cL_B &=& \fr 1 {2 \ka} (e R + \xi e B^\mu B^\nu R_{\mu\nu})
-\frac 1 4 e B_{\mu\nu} B^{\mu\nu} 
\nonumber \\
&& 
- \half e\la (B_\mu B^\mu \pm b^2)
- e B_\mu J^\mu .
\label{BBLr}
\eea
The vacuum solution for this theory 
is that of Eq.\ \rf{flatBBvac}.

The effective lagrangian $\cL_{\rm NG}$
determining the properties of the NG modes 
can be obtained by expanding the lagrangian $\cL_B$
to quadratic order,
keeping couplings to matter currents and curvature,
and using the decomposition \rf{Bprojs} of $B^\mu$.
Disregarding total derivative terms,
we find 
\bea
\cL_{\rm NG} 
&\approx &
\fr 1 {2 \ka}[
e R 
+ \xi e b^\mu b^\nu R_{\mu\nu}
+ \xi e A^\mu A^\nu R_{\mu\nu} 
\nonumber \\
&& 
+ \xi e \rh(\rh + 2) b^\mu b^\nu R_{\mu\nu} 
+ 2\xi e (\rh + 1) b^\mu A^\nu R_{\mu\nu} 
]
\nonumber \\
&&
-\frac 1 4 eF_{\mu\nu} F^{\mu\nu} - eA_\mu J^\mu 
- eb_\mu J^\mu + eb^\mu \prt_\nu \Xi_\mu J^\nu ,
\nonumber \\
\label{NGcurv}
\eea
Here,
we again relabel $A_\mu \equiv \cE_\mu$,
which obeys $b_\mu A^\mu = 0$, 
and write 
$F_{\mu\nu} \equiv \prt_\mu A_\nu - \prt_\nu A_\mu$,
which is the field strength of a gauge-fixed U(1) field.
In Eq.\ \rf{NGcurv},
the gravitational excitations $h_{\mu\nu}$
obey $h_{\mu\nu} b^\mu = 0$.
In the absence of curvature sources,
Eq.\ \rf{NGcurv}
reduces to the Minkowski-spacetime result \rf{linflatBB}.

The form of $\cL_{\rm NG}$
reveals that only two of the four potential NG modes propagate.
The propagating modes are transverse Lorentz NG modes,
while the longitudinal Lorentz and the diffeomorphism NG modes
are auxiliary.
In particular,
with the kinetic term given by the square 
of the field strength $B_{\mu\nu}$
in Eq.\ \rf{vB},
the diffeomorphism NG mode in $\rh$ again cancels,
as in the Minkowski-spacetime case.
Moreover,
the curvature terms in $\cL_B$ 
also yield no contributions for $\rh$
in $\cL_{\rm NG}$.
This is because,
in an asymptotically Minkowski spacetime,
metric excitations of the required NG form 
$h_{\mu\nu} = -\prt_\mu \Xi_\nu - \prt_\nu \Xi_\mu$
produce only a vanishing contribution 
to the curvature tensor at linear order 
and contribute only as total derivatives at second order
when contracted with $\et_{\mu\nu}$ or $b_\mu$.

As in the Minkowski-spacetime case,
it is interesting in the present context with gravity
to consider the possibility that the photon observed in nature 
can be identified with the NG mode
resulting from spontaneous Lorentz violation.
We see that,
in a Riemann spacetime,
the theory of the form \rf{NGcurv}
produces a free propagator for the Lorentz NG mode 
consistent with this idea at the linearized level.
Also,
the interaction with the charged current $J_\mu$ has
an appropriate form.
Indeed,
the effective action $\cL_{\rm NG}$ 
contains as a subset the standard Einstein-Maxwell electrodynamics 
in axial gauge.

The issue of possible experimental signals
can be revisited for the present Riemann-spacetime case.
The discussion in the previous section 
about potential SME and related signals 
in Minkowski spacetime still applies, 
but further possibilities exist.
In particular,
there are interesting unconventional couplings 
of the curvature with $A^\mu$, $\rh$, and $b^\mu$.
The photon acquires Lorentz-invariant curvature couplings
of the form $e A^\mu A^\nu R_{\mu\nu}$,
which are forbidden by gauge invariance
in conventional Einstein-Maxwell electrodynamics
but are consistent here with the axial-gauge condition.
The term $\xi e b^\mu b^\nu R_{\mu\nu}/2\ka$
corresponds to a nonzero coefficient of the $s^{\mu\nu}$ type
in the pure-gravity sector of the SME
\cite{akgrav}.
The other terms with curvature 
also represent Lorentz-violating couplings.
This lagrangian therefore gives rise to additional effects 
that could serve to provide experimental evidence 
for the idea that the photon is an NG mode 
for spontaneous Lorentz violation. 
The analysis of the associated signals is evidently of interest 
but lies outside our present scope.

It is also of interest to ask whether there exists a theory
in Riemann spacetime with a nontrivial propagator 
for the diffeomorphism mode contained in $\rh$. 
Indeed,
for a purely timelike coefficient $b_\mu$,
for which $\rh = \prt_0 \Xi^0$,
it has been shown that if a kinetic term 
for the diffeomorphism NG mode were to appear 
with second-order time derivatives,
then an unusual dispersion relation would follow
with potentially interesting phenomenological consequences
\cite{ahclt}.
In general,
the presence of curvature makes this question
more challenging than in Minkowski spacetime.

In the context of the bumblebee model \rf{BBLr}
with the field strength \rf{vB} having constant $b_a$,
we have seen that the diffeomorphism NG mode fails to propagate.
Attempting to change this by modifying the gravitational terms 
in the lagrangian \rf{BBLr} 
to any combination of covariant contractions 
of the curvature tensor $R^\ka_{\pt{\ka}\la\mu\nu}$,
including theories with general quadratic curvature terms 
\cite{svn,kf},
also fails to yield a nontrivial propagator for
the diffeomorphism NG mode
for the same reason as above.
However,
possibilities exist that might overcome this difficulty,
such as allowing for nonconstant $b_a$. 
Another interesting option is to relax the 
requirement of an asymptotically Minkowski spacetime,
perhaps by adding a cosmological-constant term to the theory.
This leads to modifications in the projector analysis
and changes in the effective action for the NG modes. 
For example, 
in a curved background a term of the form 
$\xi e \rh(\rh + 2) b^\mu b^\nu R_{\mu\nu}$
in Eq.\ \rf{NGcurv} 
would generate quadratic terms for $\rh$
in the effective lagrangian,
as needed for the propagation of $\Xi^\mu$.
A cosmological term $e \La$ also
contains quadratic terms 
$h_{\mu\nu} h^{\mu\nu} - \half h^2$ 
in the weak-field limit,
which might serve as a suitable source of quadratic terms
because a virtual diffeomorphism generates 
time derivatives for the spacelike components of $\Xi^\mu$.
Exploration of these issues is of definite interest
but lies beyond the scope of this work.

\section{Riemann-Cartan Spacetime}
\label{torsion}

In a Riemann-Cartan spacetime,
the vierbein $\vb \mu a$ 
and the spin connection 
$\nsc \mu a b$
represent independent degrees of freedom 
determined by the dynamics
\cite{rcst}.
It follows that the effects of spontaneous Lorentz breaking
can be strikingly different from the cases examined above.
In particular,
we focus in this section on the possibility 
that the NG modes are absorbed into the spin connection
via a Higgs mechanism.

\subsection{Higgs mechanism for the spin connection}

As in the Riemann spacetime case,
we suppose that a tensor 
$T_{\la\mu\nu\cdots}$ 
obeys the condition \rf{V}
and acquires a nonzero vacuum value.
This condition can be satisfied 
by expressing $T_{\la\mu\nu\cdots}$ in terms of the vierbein
according to Eq.\ \rf{vier3}.
This result can be used to calculate
the covariant derivative of the field $T_{\la\mu\nu\cdots}$,
which enters the kinetic lagrangian
for $T_{\la\mu\nu\cdots}$ 
and therefore affects the NG modes for the spontaneous Lorentz violation.
A key feature of Riemann-Cartan spacetime
is that this covariant derivative
now involves the spin connection as an independent degree of freedom.
For instance,
assuming constant $t_{\la\mu\nu\cdots}$,
the linearized expression 
in a Minkowski background is 
\beq
D_\al T_{\la\mu\nu\cdots} \approx 
\lulsc \al \rh \la  t_{\rh\mu\nu\cdots}
+ \lulsc \al \rh \mu t_{\la\rh\nu\cdots} + \ldots .
\label{Texpansion2}
\eeq
It follows that 
a kinetic term involving a quadratic expression 
in the covariant derivative of $T_{\la\mu\nu\cdots}$ 
generates a nonderivative 
quadratic expression in the spin connection.
This could represent a mass for the spin connection,
so the spontaneous violation of Lorentz symmetry
in Riemann-Cartan spacetime could
incorporate a gravitational version of the Higgs mechanism.
Note that this Higgs mechanism cannot occur in a Riemann spacetime,
where the spin connection is identically 
the derivative expression \rf{scvierb} for the vierbein,
because the same calculation produces instead 
a kinetic term for the NG modes,
as shown in the previous section. 

The lagrangian for a generic theory 
with spontaneous Lorentz violation
in Riemann-Cartan spacetime can be written as
\beq
\cL = \cL_0 
+ 
\cL_{\rm SSB},
\label{th}
\eeq
where $\cL_0$ describes the unbroken theory
and $\cL_{\rm SSB}$ induces spontaneous Lorentz violation.
We suppose for simplicity 
that $\cL_0$ contains only gravitational terms
formed from the curvature and torsion,
while the lagrangian 
$\cL_{\rm SSB}$ 
for a tensor field $T_{\la\mu\nu\cdots}$ 
contains a kinetic piece
and a potential driving the spontaneous Lorentz violation. 

{\it A priori,}
it might seem that a large range of models 
could implement this Higgs mechanism,
since numerous types of tensors
could acquire a vacuum expectation value.
However,
a complete Higgs mechanism 
requires a theory $\cL_0$
that has a massless propagating spin connection
prior to the spontaneous Lorentz violation.
A fully satisfactory example
also requires the theory to be free of ghosts.
It turns out that these conditions 
severely restrict the possibilities for model building.

General studies exist of theories 
$\cL_0$ 
with a propagating spin connection 
\cite{svn,kf},
including ones with both massive and massless propagating modes.
However,
the subset of ghost-free models is relatively small,
especially for the case of a massless spin connection.
The total number of propagating modes in these models 
depends on the presence of certain accidental symmetries.
Our investigations reveal that the symmetry-breaking lagrangian
$\cL_{\rm SSB}$ 
typically breaks one or more of the accidental symmetries of 
$\cL_0$ 
when the tensor field acquires a vacuum expectation value.
This significantly complicates the analysis of models,
but also offers interesting new avenues 
by which spontaneous Lorentz violation 
could affect the physical modes in a realistic theory.

We are primarily interested in ghost-free lagrangians $\cL_0$ 
formed from powers of the curvature and torsion
with at most two derivatives in the equations of motion
for the vierbein and spin connection.
In this case,
up to 18 of the 24 components 
of the spin connection $\nsc \mu a b$
can in principle behave as propagating degrees of freedom.
The six components $\lsc 0 a b$ are auxiliary fields,
irrespective of any gauge choices imposed for the Lorentz symmetry 
or any simplifications from accidental symmetries.

The behavior of the 18 modes can be studied
by assuming a background Minkowski spacetime
and linearizing the equations of motion
along the lines discussed in section \ref{slv}.
Note that,
at linear order in infinitesimal quantities,
the spin connection 
transforms under Lorentz transformations according to
\beq
\nsc \mu a b \rightarrow \nsc \mu a b - \prt_\mu \ep^{ab} ,
\label{omLT}
\eeq
while infinitesimal diffeomorphisms leave $\nsc \mu a b$ invariant
at lowest order.
The vacuum solution now takes the form 
\beq
\vev{T_{\mu\nu\cdots}} = t_{\mu\nu\cdots},
\quad
\vev{e_{\mu\nu}} = \et_{\mu\nu} ,
\quad
\vev{\nsc \mu a b} = 0 
\label{rcvac}
\eeq
in a suitable coordinate frame.
The vanishing of $\vev{\nsc \mu a b}$
and the invariance of $\nsc \mu a b$ under diffeomorphisms 
suggests that the fate of the diffeomorphism NG mode
is unlikely to be appreciably altered 
by the new role of the spin connection in the present context,
and this is confirmed in what follows.

\subsection{Decompositions}

The investigation of various models is facilitated
by introducing  
two different decompositions of the 24 fields $\lsc \la \mu \nu$.
The first is a decomposition according to Lorentz indices,
\beq
\lsc \la \mu \nu = 
A_{\la \mu \nu} + M_{\la \mu \nu} 
+ \frac 1 3 (\et_{\la \mu} T_\nu - \et_{\la \nu} T_\mu) ,
\label{wdecomp}
\eeq
where $A_{\la \mu \nu}$ is totally antisymmetric, 
$M_{\la \mu \nu}$ has mixed symmetry,
and $T_\mu$ is the trace piece.
The antisymmetric components define a dual 
$V_\ka = \ep_{\ka\la\mu\nu} A^{\la\mu\nu}/2$ 
that has four independent components.
The mixed components $M_{\la\mu\nu}$ satisfy eight identities,
which can be written
$M_{\nu\la}^{\pt{\nu\la}\nu} = 0$ 
and $M_{\la\mu\nu} - M_{\nu\mu\la} = M_{\mu\la\nu}$,
and they therefore contain sixteen degrees of freedom.
The trace $T_\mu$ contains the remaining four degrees of freedom.
The reader is cautioned that 
Eq.\ \rf{wdecomp} is \it not \rm a 
Lorentz-irreducible decomposition in the usual sense
because the field being decomposed is a connection
rather than a tensor.

The second useful decomposition involves 
the spin-parity projections $J^P$ 
of the fields $\lsc \la \mu \nu$.
These are particularly appropriate
for the case of timelike Lorentz violation 
induced by $t_{\mu\nu\cdots}$,
such as a bumblebee vacuum value of the form 
$b_\mu = (b,0,0,0)$. 
The 18 dynamical fields in this case include the projections 
$2^+$, $2^-$, $1^+$, $1^-$, $0^+$, $0^-$,
while the six auxiliary fields $\lsc 0 \mu \nu$ 
contain two triplet projections we denote by 
$\widetilde 1^+$, $\widetilde 1^-$.
Again, 
we stress that these projections involve the connection
rather than a tensor,
so the notation fails to reflect the true transformation properties.
For example,
the $1^+$ projection yields a triplet of scalars  
under spatial rotations.

Explicit expressions for each of these projections
can be found. 
For example,
we find
\bea
&
\om^{[0^+]} = \om_{j0j} ,
\quad
\om^{[0^-]} = \half \ep_{jkl} \om_{jkl} ,
&
\nonumber\\
&
\om_l^{[1^+]} = \ep_{jkl} \om_{j0k} ,
\quad
\om_k^{[1^-]} = \om_{jkj} ,
&
\nonumber\\
&
\om_k^{[\widetilde 1^+]} 
= \half \ep_{klm} \om_{0lm} ,
\quad
\om_k^{[\widetilde 1^-]} = \om_{00k} ,
&
\nonumber\\
&
\om_{jk}^{[2^+]} =
\frac 1 2 (\om_{j0k} + \om_{k0j}) 
- \frac  1 3 \de_{jk} \om_{l0l} ,
&
\nonumber\\
&
\om_{jk}^{[2^-]} =
\frac  1 4 (\ep_{klm} \om_{jlm} + \ep_{jlm} \om_{klm}) 
- \frac 1 6 \de_{jk} \ep_{lmn} \om_{nlm} ,
&
\nonumber\\
\label{jp}
\eea
where spatial components are denoted by $j, k, \ldots$.

The two sets of projections can be related.
We find
\bea
&
\om_k^{[1^+]}
= \om_k^{[\widetilde 1^+]} - V_k
= \fr 1 2 \ep_{k0lm} M^{0lm} - \fr 2 3 V_k ,
\label{1+}
&
\nonumber\\
&
\om_k^{[1^-]}
= - \om_k^{[\widetilde 1^-]} - T_k
= - M_{00k} - \fr 2 3 T_k ,
\label{1-}
&
\nonumber\\
&
\om^{[0^+]} = -T_0 , \quad
\om^{[0^-]} = V_0 .
\label{0+-}
\eea

\subsection{Bumblebee}

To gain further insight,
we investigate a definite form 
for the lagrangian $\cl_{\rm SSB}$,
namely,
the simple bumblebee model
\beq
\cl_{\rm SSB} = - \quar e B_{\mu\nu} B^{\mu\nu}
- e \la (B_\mu B^\mu \pm b^2) 
\label{Lssb}
\eeq
with a Lagrange-multiplier potential freezing any non-NG modes.
In a Riemann-Cartan spacetime,
the field strength $B_{\mu\nu}$ in the kinetic term 
of this theory is defined in Eq.\ \rf{Bmunu}.
Its expression in terms of the vierbein and spin connection is 
\beq
B_{\mu\nu} =
(\vb \mu b  \lulsc \nu a b - \vb \nu b  \lulsc \mu a b ) b_a .
\label{FS}
\eeq
Note that this form reduces to Eq.\ \rf{vB}
in the limits of Riemann and Minkowski spacetimes,
for which the spin connection 
is given in terms of derivatives of the vierbein 
by Eq.\ \rf{scvierb}.

When $B_{\mu\nu}$ is squared to yield the kinetic term,
quadratic terms in $\lulsc \mu a b$ appear in the lagrangian
$\cl_{\rm SSB}$.
For example,
for a Minkowski background we find
\bea
\cl_K &\equiv& - \quar e B_{\mu\nu} B^{\mu\nu} 
\nonumber\\
&\approx &
- \quar (\om_{\mu\rh\nu} - \om_{\nu\rh\mu})
(\om^{\mu\si\nu} - \om^{\nu\si\mu}) b^\rh b_\si .
\label{om2}
\eea
The appearance of these quadratic terms 
again suggests that a Higgs mechanism can occur
involving the absorption of the NG modes 
by the spin connection.

In terms of the Lorentz decomposition
in the previous subsection,
the kinetic term $\cL_K$ for $B^\mu$ becomes
\bea
\cl_{K} &\approx & 
\frac 2 9 (b_\mu b^\mu V_\nu V^\nu - b_\mu V^\mu b_\nu V^\nu)
- \frac 1 4 M^\rh_{\pt{\rh}\mu\nu} M^{\si\mu\nu} b_\rh b_\si
\nonumber\\
&&
\hskip -15pt
- \frac 1 {18} (b_\mu b^\mu T_\nu T^\nu - b_\mu T^\mu b_\nu T^\nu) 
+ \frac 1 3 \ep_{\la\mu\nu\rh} V^\rh M^{\si\la\mu} b^\nu b_\si 
\nonumber\\
&&
\hskip -15pt
- \frac 1 6 M^\la_{\pt{\la}\mu\nu} (b^\mu T^\nu - b^\nu T^\mu) b_\la .
\label{MTVSSB}
\eea
This result holds for any vacuum value $b^\mu$,
but its physical interpretation can be involved in the general case.

For the special case of timelike Lorentz violation 
induced by a vacuum expectation value $b_\mu = (b,0,0,0)$,
the $J^P$ decomposition provides a more convenient expression.
With this assumption,
we find
\beq
\cl_{\rm K} =
- \half b^2 \om_j^{[1^+]} \om^{[1^+] j} 
+ \half b^2 \om_j^{[\widetilde 1^-]} \om^{[\widetilde 1^-] j} .
\label{SSB3}
\eeq
We see that this expression
contains an apparent physical mass term for the $1^+$ 
and a wrong-sign mass term for the $\widetilde 1^-$.
Since the $\widetilde 1^-$ is an auxiliary field,
it cannot propagate independently.
However,
the $1^+$ is an independent dynamical field,
so interpreting its apparent mass term
requires a study of the dynamical content of $\cL_0$.

\subsection{Illustrative models}

Next,
we present three different sample models $\cL_0$,
all containing dynamical terms for the spin connection,
to illustrate some of the possible effects and issues 
emerging from the presence of the Lorentz-breaking term 
$\cl_{\rm SSB}$.
In the first model,
denoted $\cl_{0,1}$,
ghosts are present
but an analysis shows that a Higgs mechanism occurs
when $\cl_{\rm SSB}$ is added.
The second,
$\cl_{0,2}$,
initially has only auxiliary or gauge degrees of freedom,
but the addition of the Lorentz-violating term $\cl_{\rm SSB}$
breaks some accidental symmetries
and hence causes some modes to propagate.
The third model,
$\cl_{0,3}$,
is ghost free and has a massless propagating spin connection.

The lagrangian for the first example is
\beq
\cl_{0,1} =  
\frac 1 4 R_{\la \ka \mu \nu} R^{\la \ka \mu \nu} .
\label{R2L}
\eeq
To lowest order in the spin connection,
the curvature tensor becomes 
$R_{\la \ka \mu \nu} \approx
\prt_\ka \lsc \la \mu \nu - \prt_\la \lsc \ka \mu \nu$.
In this model,
all the fields $\lsc \la \mu \nu$ with $\la \ne 0$
propagate as massless modes.
However,
when resolved into $J^P$ projections,
the second-derivative terms in the equations of motion for
the even- and odd-parity states have opposite signs,
so the theory contains ghosts.
When $\cl_{0,1}$ is combined with $\cl_{\rm SSB}$,
the linearized equations of motion become
\bea
\hskip -10pt
\prt_\rh \prt^\rh \lsc \la \mu \nu
- \prt_\la \prt^\rh \lsc \rh \mu \nu
&=&
- \half ( \lsc \la \si \nu - \lsc \nu \si \la) b_\mu b^\si
\nonumber\\
&&
+ \half ( \lsc \la \si \mu - \lsc \mu \si \la) b_\nu b^\si .
\label{one}
\eea
These 24 equations can be diagonalized to determine
the nature of the modes in the combined theory,
and we find that among the propagating modes is a massive one.
This confirms the existence of a Higgs mechanism 
for the spin connection in this model.

The idea behind the second model is to start with
a special theory $\cL_0$ in which accidental symmetries
exclude all propagating physical modes,
but chosen such that physical propagating modes emerge
when the lagrangian $\cl_{\rm SSB}$ 
triggering spontaneous Lorentz violation is added.
The appearance of the physical modes via this  
`phoenix' mechanism 
can be traced to the breaking of some accidental symmetries 
of $\cL_0$ by $\cl_{\rm SSB}$.

A number of models in which all modes 
are auxiliary or gauge are known
\cite{kf}.
Here,
we consider one explicit example,
with lagrangian given by
\beq
\cl_{0,2} = 
\half R_{\mu\nu} R^{\mu\nu} - \half R_{\mu\nu} R^{\nu\mu} ,
\label{unbrok}
\eeq
where $R_{\mu\nu}$ is the Ricci tensor in Riemann-Cartan spacetime.
Since our focus is on the spin connection,
we restrict attention for simplicity 
to solutions in background Minkowski spacetime. 
With this choice,
the vierbein disappears from the linearized theory,
so the spin connection is the only relevant dynamical field.

The unbroken lagrangian can be written in terms of the
Lorentz decomposed fields as
\bea
\cl_{0,2} &=& 
\frac 1 9 F_{\mu\nu} F^{\mu\nu} 
- \frac 1 9 G_{\mu\nu} G^{\mu\nu}
+ \frac 1 4 \prt_\rh M^\rh_{\pt{\rh}\mu\nu} \prt_\si M^{\si\mu\nu} 
\nonumber\\
&& 
- \frac 1 3 (F_{\mu\nu} + \half \ep_{\mu\nu\rh\si} G^{\rh\si}) 
\prt_\la M^{\la\mu\nu}.
\label{MTVfree}
\eea
The corresponding equations of motion are 
\bea
&
\prt_\la [ \prt^\si M_{\si\mu\nu}
- \fr 2 3 (F_{\mu\nu} 
+ \half \ep_{\mu\nu\rh\si} G^{\rh\si}) ] = 0 ,
\label{Meq}
&
\\
&
\prt^\mu F_{\mu\nu} 
= \fr 3 2 \prt^\mu \prt^\si M_{\si\mu\nu} ,
\label{Teq}
&
\\
&
\prt^\mu G_{\mu\nu} = 
- \fr 3 4 \ep_{\mu\nu\si\rh} \prt^\si \prt^\la M_\la^{\pt{\la}\rh\mu} .
\label{Veq}
\eea
In these equations,
$F_{\mu\nu} = \prt_\mu T_\nu - \prt_\nu T_\mu$
and
$G_{\mu\nu} = \prt_\mu V_\nu - \prt_\nu V_\mu$ 
are the field strengths
for $T_\mu$ and $V_\nu$, respectively.

A cursory inspection might suggest 
that this theory has at least 
two sets of massless fields, 
$T_\mu$ and $V_\nu$,
which correspond to the $1^+$ and the $1^-$ modes 
in the $J^P$ decomposition.
However,
there are a number of accidental symmetries in
this theory associated with the
projection operators for the $2^+$, $2^-$, $0^+$, and $0^-$ fields.
These and Lorentz transformations can be used
to remove all physical propagating degrees of freedom
\cite{kf}.
In particular,
the $1^-$ mode can be gauged away using only rotations,
while the $1^+$ mode can be gauged away using only boosts:
\bea
\om_j^{[1^-]} &\rightarrow &
\om_j^{[1^-]} - \prt^k \ve_{jk}
\nonumber\\
\om_j^{[1^+]} &\rightarrow &
\om_j^{[1^+]} + \ep_{j0}^{\pt{j0}lm} \prt_l \ve_{0m}.
\label{oneplusminus}
\eea
The net result is that the Lorentz-invariant theory \rf{unbrok}
has no physical content.

Suppose now the term $\cl_{\rm SSB}$ in Eq.\ \rf{SSB3}
for the case of a timelike vacuum expectation value $b_\mu$ 
is added to the lagrangian \rf{unbrok}.
This spontaneously breaks boosts 
while maintaining rotation symmetry.
The $1^-$ mode can still be gauged away via rotations,
but the $1^+$ mode can no longer be removed using boosts
and so might be expected to propagate as a massive mode.
However,
the mass term in $\cl_{\rm SSB}$ 
also affects the structure 
of the gauge and auxiliary fields in the theory
by breaking some of the accidental symmetries,
so other field combinations now become physical.
We find two such massless modes,
involving superpositions of the $J^P$ projections.

For our third example,
we take for $\cL_0$ a ghost-free model
with a massless propagating spin connection.
A general analysis 
under the assumption of Lorentz invariance
finds only four ghost-free possibilities 
\cite{kf}.
All share the property of the previous example
that the propagating modes 
consist of mixtures of $J^P$ projections.
In two models,
the massless propagating mode incorporates 
contributions from the $1^+$ projections,
while in the other two it includes
contributions from the $1^-$ projections.
It is therefore of interest to adopt for $\cL_0$
either of the first two models
and investigate the effect
on the propagating modes of adding the Lorentz-violating term 
$\cl_{\rm SSB}$.

As an explicit example,
consider the lagrangian
\cite{kf}
\beq
\cl_{0,3} = R_{\mu\nu} R^{\mu\nu} - \frac 1 3  R^2 .
\label{unbrok3}
\eeq
It turns out that the propagating modes 
of this Lorentz-invariant model are a mixture of 
$1^+$ and $2^+$ projections.
In contrast,
$\cl_{\rm SSB}$
contains quadratic terms for the $1^+$
and the auxiliary $\tilde 1^-$ states.
In the theory resulting from the combination of the two,
the nature of the modes can be determined by diagonalizing
the 24 linearized equations for the spin connection.
We find that the propagation of the massless modes is altered,
but there is no massive propagating $1^+$.
The incompatibility between the mixture of $J^P$ states
appearing in $\cl_{0,3}$ and that appearing in $\cl_{\rm SSB}$
prevents the occurrence of a clean Higgs mechanism 
for the $J^P$ modes in this example.
 
In the context of these ideas,
a number of issues of interest 
remain open for future investigation.
Studies of the large variety of possible 
Lorentz-invariant lagrangians $\cL_0$
could lead to additional features beyond those 
identified in the three examples above.
It would also be of interest to explore 
more explicitly the effects of lightlike and spacelike $b_\mu$
in the Riemann-Cartan spacetimes.
The $J^P$ decomposition is less appropriate for these cases,
so alternative decompositions with respect to the 
corresponding little group are likely to be useful. 
Different choices for $\cl_{\rm SSB}$,
including ones in which the spontaneous Lorentz violation
involves one or more tensor fields,
can also be expected to affect the dynamics of the NG modes.
From a broader theoretical perspective,
the incorporation of Lorentz violation opens
an arena for the search for  
ghost-free theories with dynamical curvature and torsion.

Various implications for phenomenology in the context of
Riemann-Cartan spacetime also merit exploration.
The scale of the emergent mass in the models considered here
is set by $b^2$.
Even if this is of order of the Planck mass,
the existence of fields with Lorentz-violating physics
could have effects on cosmology
and in regions with strong gravitational fields
such as black holes.
The couplings to other known fields also merit attention
and could lead to interesting signals for experiments.
All relevant terms associated 
with gravitational and SM fields
are included in the gravitational couplings 
of the Lorentz- and CPT-violating SME
in Riemann-Cartan spacetime
\cite{akgrav},
which therefore provides the appropriate framework for
investigating phenomenological implications of these models.

\section{Summary}
\label{concl}

In this paper,
we have examined the fate of the Nambu-Goldstone modes 
when Lorentz symmetry is spontaneously broken.
The analysis is performed in the context 
of the vierbein formalism,
which is well suited for this purpose
because it admits a clear separation  
between local Lorentz and coordinate frames 
on the spacetime manifold.
Within this formalism,
we have demonstrated in section \ref{slv}
that spontaneous particle Lorentz violation 
is accompanied by spontaneous particle diffeomorphism violation
and vice versa,
and that up to 10 NG modes can appear.
These modes can naturally be matched 
to those 10 of the 16 modes of the vierbein
that in a Lorentz-invariant theory are 
gauge degrees of freedom.
This match provides further evidence
for the value of the vierbein formalism
in studies of spontaneous violations of spacetime symmetries.
We have also provided a generic treatment 
for background Minkowski spacetimes.
The fate of the NG modes
is found to depend both on the spacetime geometry 
and also on the dynamics of the tensor field triggering 
the spontaneous violation 
of local Lorentz and diffeomorphism symmetries.

As illustrative models for the analysis,
we have adopted a general class of bumblebee models,
involving vacuum values for a vector field
that break some of the local Lorentz and diffeomorphism symmetries.
Some properties of these models have been presented in 
section \ref{bbms},
where projectors are constructed that permit separation 
of the Lorentz and diffeomorphism NG modes.

In the later sections of this work,
we have studied the behavior of the NG modes
in Minkowski,
Riemann,
and Riemann-Cartan spacetimes.
Each of these offers distinctive general features,
which can be illustrated within bumblebee models.
In Minkowski and Riemann spacetimes,
Lorentz NG modes exist that can propagate as massless modes,
with effective lagrangians containing
the Maxwell and Einstein-Maxwell theories in axial gauge.
Suitable bumblebee models thereby provide dynamical methods 
of generating the photon as a Nambu-Goldstone boson  
for spontaneous Lorentz violation.
Various possibilities exist for experimental signals
in these models,
including both unconventional Lorentz-invariant couplings
and Lorentz-breaking couplings 
in the matter and gravitational sectors
of the SME.
In Riemann-Cartan spacetimes,
the interesting possibility exists 
that the spin connection could absorb the propagating NG modes 
in a gravitational version of the Higgs mechanism.
This unique feature of gravity theories with torsion
may offer another phenomenologically viable route 
for constructing realistic models
with spontaneous Lorentz violation. 

\section*{Acknowledgments}

This work was supported in part
by the Department of Energy
under grant number DE-FG02-91ER40661,
by the National Aeronautics and Space Administration
under grant numbers NAG8-1770 and NAG3-2194,
and by the National Science Foundation
under grant number PHY-0097982.


\begin{thebibliography}{99}
\bibitem{note2}
Under special circumstances more than 10 generators are involved.
For example,
in a theory based on special relativity
with multiple noninteracting particle species,
there are distinct particle Lorentz transformations 
for each field.
The particle symmetry group of the theory then 
contains a direct product of several Lorentz groups, 
and so there could be multiple sets of NG modes.
However, 
in typical gravitational theories 
with the metric coupling to all irreducible tensors,
the full particle invariance involves only one copy 
of the particle Lorentz and diffeomorphism groups, 
so only ten corresponding NG modes can appear.
\bibitem{note3}
Note that,
in a Lorentz-invariant theory only,
the effects of the particle transformations
considered in this subsection
are equivalent in their action on tensor components
to certain inverse observer transformations.
The Minkowski-spacetime transformations \rf{plt}
then correspond to observer Lorentz transformations 
to a local frame with an anholonomic basis,
and the diffeomorphisms leading to Eq.\ \rf{metricmk}
correspond to working with special relativity 
in a non-cartesian basis.
\end{thebibliography}
\end{document}